\documentstyle[11pt,aaspp4]{article}
\def\N{N}

\def\gsim{\ga}

\begin{document}

\title{Constraints on Galaxy Evolution and the Cosmological \\ Constant From
Damped Ly$\alpha$ Absorbers}
\author{Eric Woods$^1$ and Abraham Loeb$^2$}
\medskip
\affil{Astronomy Department, Harvard University, 60 Garden St.,
Cambridge, MA 02138}
\altaffiltext{1}{email: ewoods@cfa.harvard.edu}
\altaffiltext{2}{email: aloeb@cfa.harvard.edu}

\begin{abstract}

We use the existing catalog of Damped Lyman--Alpha (DLA) systems to place
constraints on the amount of evolution in the baryonic content of galaxies
and on the value of the cosmological constant.  The density of cold gas at
redshifts $z=3\pm 1$ is obtained from the mean HI column density of DLAs
per cosmological path length.  This path length per unit redshift is in
turn a sensitive function of the vacuum density parameter,
$\Omega_\Lambda$.  We compare the total inferred mass of cold gas at high
redshifts to that observed in stars today for cosmologies with $\Omega_{\rm
m}+\Omega_\Lambda=1$, where $\Omega_{\rm m}$ is the matter density
parameter.  We define $\eta$ to be net fraction of the baryonic content of
local galaxies which was expelled since $z=3$, and use Bayesian inference
to derive confidence regions in the ($\eta, \Omega_\Lambda$) plane.  In all
cosmologies we find that $\eta<0.4$ with at least $95\%$ confidence if
$<25\%$ of the current stellar population formed before $z=3$. The most
likely value of $\eta$ is negative, implying a net {\it increase} by
several tens of percent in the baryonic mass of galaxies since $z=3\pm 1$.
On the other hand, recent observations of high metal abundances in the
intracluster medium of rich clusters (Loewenstein \& Mushotzky 1996)
require that metal--rich gas be {\it expelled} from galaxies in an amount
approximately equal to the current mass in stars.
Based on our results and the low metallicity observed in DLAs at $z\ga2$,
we infer that more than half of the baryonic mass processed through
galaxies must have been assembled and partly expelled from galaxies {\it
after} $z=2$.  We expect our constraints to improve considerably as the
size of the DLA sample will increase with the forthcoming Sloan Digital Sky
Survey.

\end{abstract}

\centerline{Submitted to {\it The Astrophysical Journal}, March 1997}

\vfill\eject

\section{Introduction}

To date, some $\sim 80$ Damped Lyman-Alpha (DLA) absorption systems have
been identified in the spectra of high-redshift QSOs.  The observed
absorption troughs indicate concentrations of neutral gas, with large HI
column densities, $N\ga 10^{20}$ cm$^{-2}$.  There is considerable evidence
that these objects are associated with the progenitors of present--day
galaxies.  This has been confirmed in several cases by direct imaging of
QSO fields (Steidel et al. 1994, 1995, 1996; Djorgovski et al. 1996; Le
Brun et al. 1996).  There are various indications that DLAs are associated
with young galaxies.  The abundances of metals at low ionization stages in
DLAs are comparable to those found in disk galaxies (Wolfe 1995).  The
metal absorption line profiles show a leading--edge asymmetry (Wolfe 1995;
Prochaska \& Wolfe 1997) and are shifted relative to the Ly$\alpha$
emission redshift (Lu, Sargent, \& Barlow 1997), as expected from a
rotating thick disk with circular velocities comparable to those seen in
spiral galaxies.  Furthermore, observations of redshifted 21-cm absorption
and emission from DLAs indicate disk-like structures of galactic dimensions
(Briggs et al.  1989; Wolfe et al. 1992).  However, recent HST images (Le
Brun et al. 1996) have revealed that DLA galaxies span a wide range of
morphological types.  The existence of a substantial galaxy population at
redshifts $z\gsim 2$ is consistent with most CDM models, which predict that
galaxies should form by $z\approx 2-3$ (Frenk et al.  1996, and references
therein).

The HI column densities in DLAs can be derived from the equivalent widths
of the observed absorption features.  The mean HI column density along a
random line of sight may then be summed up and divided by the absorption
path length probed to obtain the comoving spatial HI mass density
$\rho_{\rm HI}$ in galaxies at a given redshift.  Since DLAs dominate the
HI mass in the universe (Lanzetta et al. 1995, and references therein),
their statistics can be used to infer the evolution of the comoving HI
density from high redshifts up to the present epoch.  Recent work (Lanzetta
et al. 1995; Wolfe et al. 1995; Storrie-Lombardi, McMahon, \& Irwin 1996)
has shown that, for universes with a zero cosmological constant
($\Lambda=0$), the inferred comoving total gas density $\rho_{\rm
g}=1.3\rho_{\rm HI}$ (including HI and He) at a redshift $z\approx 3.5$ is
comparable to the stellar mass density observed in present--day galaxies.
This was seen to be consistent with a simple ``closed--box'' picture in
which galaxies had formed by $z=3.5$, and the neutral gas was subsequently
converted into stars, while the total baryonic mass (gas + stars) is
conserved.  Note that the closed-box model does not assume that the
baryonic masses of individual galaxies are conserved but rather that the
total baryonic mass of {\it all} galaxies is not evolving with time. This
model allows for mergers which conserve the total mass of all of the
galaxies involved.

Evolutionary models may be generalized to include a net infall of gas from
the intergalactic medium, or a net outflow of gas from DLA galaxies.  The
closed-box assumption represents the simplest evolutionary model, but in
general we do not expect the total baryonic mass in galaxies to be
conserved.  In particular, the intracluster medium (ICM) in rich clusters
of galaxies contains an amount of iron which is approximately equal to the
amount locked in stars in these clusters (Renzini et al. 1993; Loewenstein
\& Mushotzky 1996).  The natural explanation is that the metals observed in 
the ICM
were produced inside of galaxies and subsequently expelled into the ICM by
supernova--driven winds.  In this paper, we examine whether the DLA
data are consistent with the substantial outflow needed to account for this
observation.

In the subsequent discussion, we also allow $\Lambda\neq 0$, and explore the
family of models with
$\Omega_{\rm m}+\Omega_\Lambda = 1$, where $\Omega_{\rm m}$ and
$\Omega_\Lambda$ are the mean present--day cosmic mass densities in matter
and vacuum energy, respectively, in units of the critical density.  
The absorption path length corresponding to a given redshift interval
depends on the assumed cosmological model; it is shortest for flat,
$\Lambda=0$ universes, longer for open, $\Lambda=0$ universes, and longer
still for flat, $\Lambda\neq 0$ models.  Because the absorption path length
is longest for $\Lambda$--dominated cosmologies, we expect the inferred
value of $\rho_{\rm g}$ at high redshifts to be the smallest, and hence
the deficit with respect to the present--day stellar density to be the
largest in these cosmologies.  This approach is similar to methods which
constrain $\Omega_\Lambda$ based on the statistics of gravitational lensing 
(Kochanek 1996, and references therein).
%Fukugita \& Turner 1991; Fukugita et al. 1992; 
%Kochanek 1992, 1996; Rix et al. 1994).
In closed--box models, one may investigate the constraints that can be
placed on $\Omega_\Lambda$ by requiring, for some suitably chosen $z>2$,
that $\rho_{\rm g}(z)$ be comparable to the present-day mass density in
stars, $\rho_{\rm s}(0)$.  However, a net infall of gas between the
high-redshift epoch and $z=0$ can bring the DLA data into reasonable
agreement with high--$\Lambda$ cosmologies.  For our purposes, it is
sufficient to parameterize the amount of accretion or expulsion as
$\eta\equiv \rho_{\rm b,gal}(z)/\rho_{\rm b,gal}(0) - 1$, where $\rho_{\rm
b,gal}=\rho_{\rm g}+\rho_{\rm s}$ is the total baryonic density in
galaxies, and $\rho_{\rm b,gal}(0)\approx\rho_{\rm s}(0)$.  In this paper
we explore the constraints on both galaxy evolution and cosmology by
constructing confidence regions in the $(\eta,\Omega_\Lambda)$ plane.  To
simplify our analysis, we make the reasonable assumption that the
distribution of DLAs in HI column density $N$ and redshift $z$, $F(N,z)$,
is separable into functions of $N$ and $z$ over a suitably small redshift
interval. 
This allows us to fit the column density distribution separately from the
extraction of constraints on galaxy formation or the underlying cosmology
[the latter being exclusively related to the redshift dependence of
$F(N,z)$].

The value of $\rho_{\rm s}(0)$ is calculated by multiplying the local
luminosity density by the mean mass-to-light ratio.  Recent imaging of DLA
galaxies (Le Brun et al. 1996) indicates that, while some are spirals, a
significant fraction of these objects have irregular morphologies, which
may indicate that a substantial amount of merging was taking place at high
redshifts.  Since mergers may result in the formation of elliptical
galaxies, we include all galaxy types in our calculation of the local
luminosity density, and hence $\rho_{\rm s}(0)$. The assumption that
underlies our discussion is that star formation requires {\it cold} HI gas,
which must be represented in a fair sample of all Ly$\alpha$ absorption
systems, irrespective of whether the star formation process occurs in
spirals or in ellipticals.

In \S 2.1 we show how the comoving HI density is inferred from the DLA
sample, and how sensitive it is to the underlying cosmology.
Section 2.2 adds the impact that evolution in the HI content
of galaxies might have on our analysis.  In \S 3 we discuss the statistical
methods used to compare the data to the theoretical predictions for
$\rho_{\rm g}(z)$.  In \S 4 we present the derived confidence intervals
for our constraints on the evolution of galaxies and the cosmological
constant.  Finally, \S 5 summarizes our conclusions.  

\section{Evolution Of The Comoving HI Density}

In order to extract useful constraints from the data, we must predict some
observable property of the DLA sample, and show explicitly how our prediction
depends on our assumptions about cosmology and evolution.  In \S 2.1, we
show how, for a given cosmology, the comoving HI density $\rho_{\rm HI}(z)$
is related to the total DLA column density along a line of sight.  In \S 2.2,
we introduce a parameter which characterizes the amount of evolution in the
HI density.

\subsection{Inferring $\rho_{\rm HI}(z)$ from the DLA Sample:
Effects of Cosmology}

The comoving HI density in DLA systems at a redshift $z$ is inferred by
calculating the mean HI column density in a proper length interval
$cdt$ along a line of sight, and dividing the result by $(c/H_0)dX\equiv
(1+z)^3cdt$.  Since the absorption path length element $dX$ corresponding
to a given redshift element $dz$ depends on the cosmological parameters
$\Omega_{\rm m},\Omega_\Lambda$ (the mean present-day cosmic mass densities
in matter and vacuum energy, respectively, in units of the critical
density), the inferred value of the comoving HI density will depend on the
assumed geometry of the universe.  Let $F(N,z)dNdz$ be the mean number of
DLAs along a line of sight with HI column densities between $N$ and $N+dN$
and redshifts between $z$ and $z+dz$.  Note that $F(N,z)$ is different from the
function $f(N,z)$ usually encountered in the literature; the latter is conventionally
defined such that $f(N,z)dNdX$ gives the number of DLAs with with column densities
between $N$ and $N+dN$ and absorption distances between $X(z)$ and $X(z)+(dX/dz)dz$.
The inferred comoving HI density
is given by
\begin{equation}
\rho_{\rm HI}(z) = \left({H_0m_{\rm H}\over c}\right)
\left[{{dz}\over{dX}}(z,\Omega_m,\Omega_\Lambda)\right] \int_{N_{\rm min}}^
\infty F(N,z)NdN,
\label{eq:rhomean}
\end{equation}
where $m_{\rm H}$ is the mass of a hydrogen atom, and $N_{\rm min}=2\times 10^{20}$
cm$^{-2}$ is the minimum HI column density included in the sample.  For a
cosmological model with density parameters ($\Omega_{\rm m},\Omega_\Lambda$) 
and an open or flat
geometry, the absorption path length element $dX$ is given by 
\begin{equation}
dX =
{{(1+z)^2dz}\over{\sqrt{\Omega_{\rm m}(1+z)^3 + (1-\Omega_{\rm m}-
\Omega_\Lambda)(1+z)^2 + \Omega_\Lambda}}}.
\label{eq:dxdz}
\end{equation}

For a flat matter--dominated universe with $\Omega_{\rm m}=1$ and
$\Omega_\Lambda=0$, $dX=(1+z)^{1/2}dz$; for a low--density universe with
$\Omega_{\rm m}=0$ and $\Omega_\Lambda=0$, $dX=(1+z)dz$; and for a flat,
$\Lambda$--dominated universe with $\Omega_{\rm m}=0$ and
$\Omega_\Lambda=1$, $dX=(1+z)^2dz$.  Hence, because the path
length corresponding to a given redshift interval $dz$ is longest for
$\Lambda$--dominated cosmologies, the $\Omega_{\rm HI}(z)$ value inferred
from equation~(\ref{eq:rhomean}) is smallest for large $\Omega_\Lambda$.  In this
paper we will only consider the family of flat cosmologies with
$\Omega_{\rm m}+\Omega_\Lambda=1$, and will write all cosmology--dependent
expressions in terms of the single adjustable parameter $\Omega_\Lambda$.

The {\it observed} distribution $F_{\rm obs}(N,z)dNdz$, defined as the
{\it total} number of DLAs observed in the sample with column densities
between $N$ and $N+dN$ and redshifts between $z$ and $z+dz$, depends on the
sensitivity of the QSO sample to the detection of a DLA feature at redshift
$z$.  In the absence of obscuration of background QSOs by dust
in the DLAs,
\begin{equation}
F_{\rm obs}(N,z) = g(z)F(N,z),
\label{eq:fobs}
\end{equation}
where
\begin{equation}
g(z) \equiv \sum_{i=1}^m H(z_i^{\rm max}-z) H(z-z_i^{\rm min}).
\label{eq:gz}
\end{equation}
Here, $H(x)$ the Heaviside step function, $m$ is the total number of
QSOs in the sample, and $(z_i^{\rm min},z_i^{\rm max})$ is the redshift
window over which the observations are able to detect a DLA feature in the
spectrum of the $i$th QSO (depending on the redshift of the QSO and the
response of the detector).  Thus, $g(z)$ is the number of lines of sight
for which a DLA feature is detectable at an absorber redshift $z$.  We
therefore have
\begin{equation}
\rho_{\rm HI}(z) = \left({H_0m_{\rm H}\over c}\right)
\left[{{\sqrt{(1-\Omega_\Lambda)(1+z)^3 + \Omega_\Lambda}}\over{(1+z)^2}}\right]
{1\over{g(z)}} \int_{N_{\rm min}}^\infty F_{\rm obs}(N,z)NdN.
\label{eq:rhotot}
\end{equation}
Equation~(\ref{eq:rhotot}) expresses the inferred value of $\rho_{\rm HI}(z)$,
factored into functions that depend exclusively on $\Omega_\Lambda$, which we
treat as a free parameter, and the properties of the QSO and DLA samples,
which are fixed by observations.

\subsection{Predicting $\rho_{\rm HI}(z)$: Effects of Evolution}

Given a specified galaxy evolution picture, the evolution of the gas
density is obtained by solving the equations
of cosmic chemical evolution (e.g. Pei \& Fall 1995):
\begin{equation}
\dot{\rho}_{\rm g} + \dot{\rho}_{\rm s} = \dot{\rho}_{\rm b,gal},
\label{eq:masscons}
\end{equation}
\begin{equation}
\rho_{\rm g}\dot{Z} - y\dot{\rho}_{\rm s} = (Z_{\rm f}-Z)\dot{\rho}_{\rm b,gal}.
\label{eq:metalev}
\end{equation}
Here, the dot denotes a time derivative; $\rho_{\rm g}$ is the total gas
density (note that $\rho_{\rm g}>\rho_{\rm HI}$, since $\rho_{\rm g}$
includes HII, H$_2$, He, and heavier elements); $\rho_{\rm s}$ is the mass
density in stars; $\dot{\rho}_{\rm b,gal}$ is the net rate at which the
total baryonic mass density changes ($\dot{\rho}_{\rm b,gal}>0$ corresponds
to a net accretion of material from the intergalactic medium, while
$\dot{\rho}_{\rm b,gal}<0$ corresponds to a net expulsion of material from
galaxies); $y$ is the mean stellar yield (mass fraction of elements heavier
than He produced in stars), averaged over the
stellar initial mass function (IMF); $Z$ is the metallicity of the gas
(mass fraction of elements heavier than He present in the gas) in galaxies,
and $Z_{\rm f}$ is the metallicity of the infalling gas.  The simplest
solution is the ``closed-box'' solution, which has no net accretion or
outflow ($\dot{\rho}_{\rm b,gal}=0$):
\begin{equation}
\rho_{\rm g}(z) = \rho_{\rm g}(\infty) \exp\left[-{{Z(z)}\over y},
\right]
\label{eq:closedbox}
\end{equation}
where we have assumed $Z(\infty)=0$.  Solutions which include accretion or
outflow have been identified by Pei \& Fall (1995) for the case where
$\dot{\rho}_{\rm b,gal}\propto\dot{\rho}_{\rm s}$.

For simplicity, we will assume that $\rho_{\rm g}$ is constant within a
sufficiently narrow redshift interval around $z=3$.  Although the minimum
width of this interval is limited to $\Delta z\sim$1--2 by the current size
of the DLA sample, one can imagine that forthcoming DLA surveys will allow
one to narrow it much more in the future.  In principle, one could have
considered the case of a constant nonzero rate of change $\dot\rho_{\rm
g}$; however, even with $\Delta z\sim$1--2 this produces results which do
not differ appreciably from the case where $\rho_{\rm g}$ is a constant,
for a range of reasonable values of $\dot\rho_{\rm g}$.

Any deficit in the gas density $\rho_{\rm g}(z)$ at $z=3$ relative
to the total present--day baryonic mass $\rho_{\rm b,gal}(0)=\rho_{\rm
g}(0)+\rho_{\rm s}(0)\approx\rho_{\rm s}(0)$ will be due to a combination
of two effects: (1) some material may have accreted onto existing galaxies
or assembled into new galaxies since $z=3$, and (2) some star
formation may have already occurred by redshift $z=3$, depleting part of
the gas.  Effect (2) is expected to be sub--dominant based on the low
metallicities $Z\sim 0.1Z_\odot$ observed at redshifts $z\gsim 2$ (Lu
et al. 1996) and the relatively small star formation rates
observed at such high redshifts (Madau 1996).  Nevertheless, we will lump
the two effects together, and will parameterize the combined contribution
to the deficit by defining
\begin{equation}
\eta\equiv {{\rho_{\rm b,gal}(3)}\over{\rho_{\rm b,gal}(0)}} - 1,
\label{eq:etadef}
\end{equation}
where $1+\eta$ is the fraction of $\rho_{\rm b,gal}(0)$ which was present
at a redshift of 3.  (We choose $z=3$ as the fiducial redshift for the
comparison, but since we assume that $\rho_{\rm g}$=constant over an
interval containing $z=3$, we could just as easily choose any other
redshift in this interval for the definition of $\eta$.)  Hence, $\eta<0$
corresponds to a net accretion and $\eta>0$ corresponds to a net expulsion
of material since $z=3$.  Star formation prior to $z=3$ is expected to
produce at most $\sim$25\% of $\rho_{\rm s}(0)$ (approximately the
fractional area under the the curve $\dot{\rho}_{\rm s}(t)$ inferred from
Madau 1996, corrected for high-$\Lambda$ cosmologies).  Let $f$ be the
fraction of the total present-day mass in stars which were produced by
$z=3\pm 1$; then, with $\rho_{\rm b,gal}(0)\approx\rho_{\rm s}(0)$, we have
\begin{equation}
\rho_{\rm g}(3) = (1 + \eta - f) \rho_{\rm s}(0).
\label{eq:rhog}
\end{equation}

Finally, we neglect the contributions of ionized hydrogen, molecular hydrogen
(see Ge \& Bechtold 1997), as well as metals, to $\rho_{\rm g}$ at
redshifts $z\gsim 2$, but include helium, 25\% by mass, resulting in a mean
molecular weight of 1.3$m_{\rm H}$.  We therefore use
$\rho_{\rm g} = 1.3\rho_{\rm HI}$.

\section{Statistical Methods}

Next we describe the methods used to compare the data to the model
prediction $\rho_{\rm HI}(z)$ for the mean HI density at high redshift, at
given values of $\eta$ and $\Omega_\Lambda$.  In \S3.1, we construct the
likelihood function, and in \S3.2, we use Bayesian analysis to derive
confidence regions in the ($\eta,\Omega_\Lambda$) plane from the likelihood
function.

\subsection{Constructing a Likelihood Function}

The likelihood function gives the probability of obtaining a particular
data set, assuming the truth of a specified model.  The model should
include predictions for both the expected value of the quantity in question
and the {\it distribution} of measurements due to statistical (e.g.
Poisson) fluctuations about the expected value.  The most commonly used
likelihood estimator in the literature is the $\chi^2$ statistic, which is
simply proportional to the logarithm of the likelihood function for
normally distributed errors.  Its widespread use is due to two convenient
facts: the tendency, due to the Central Limit Theorem, of sums of many
random variables to be normally distributed; and the fact that the
distribution of the $\chi^2$ statistic for normally distributed errors is
known, which makes it easy to calculate confidence regions.  However, the
$\chi^2$ statistic is not appropriate for our analysis here.  To apply the
$\chi^2$ statistic, we would need to collect the data into redshift bins,
and estimate the value of $\rho_{\rm HI}$ in each bin by summing up the
column densities and dividing by the total absorption path length probed in
the bin.  However, given our small sample size (only 73 objects in total),
such experimentally-measured values of $\rho_{\rm HI}$ will not be normally
distributed.
In fact, the values we measured from Monte-Carlo-simulated data sets show a
significant skewness in their distribution, and so we are not justified in
using the $\chi^2$ statistic to calculate confidence regions. 
Moreover, this approach introduces arbitrariness in the choice of bin size
and also loses information about the DLA sample by binning the data.  We
have therefore chosen to construct a likelihood function which makes use of
the {\it unbinned} data, and which does not rely on the assumption of
normally distributed errors.

The starting point for constructing our likelihood function is to note that
any prediction for $\rho_{\rm HI}(z)$ can be related to the observed
distribution of DLAs in column density and redshift using equation~(\ref{eq:rhotot}). 
Since we assume that $\rho_{\rm HI}(z)$ is approximately
constant over some narrow redshift interval around $z=3$, we
may use equation~(\ref{eq:rhog}), and rearrange ~(\ref{eq:rhotot}) to write 
\begin{equation}
\int_{N_{\rm min}}^\infty F_{\rm obs}(N,z)NdN = 
\left({c\over{H_0m_{\rm H}}}\right)
\left[{{(1+z)^2g(z)}}\over{\sqrt{(1-\Omega_\Lambda)(1+z)^3 + \Omega_\Lambda}}\right]
{\ }\rho_{\rm s0}(1+\eta-f),
\label{eq:fndn}
\end{equation}
where $\rho_{\rm s0}\equiv \rho_{\rm s}(0)$.  Thus, given an evolution
model $\rho_{\rm HI}(z)$ and a cosmological model $\Omega_\Lambda$, we have
a definite prediction for a property of the DLA sample, namely the total
column density as a function of redshift [given by the left hand side of
equation~(\ref{eq:fndn})].  Note that~(\ref{eq:fndn}) gives the {\it expected} value of
this quantity; it is not obvious how the values measured from hypothetical
data sets should be distributed about this expected value (as mentioned
above, given the small sample size, the values in each redshift bin will
not be normally distributed).  Specifically, the distribution of measured
$\rho_{\rm HI}(z)$ values depends on the shape of $F_{\rm obs}(N,z)$ as a
function of HI column density $N$, and will be skewed toward smaller values
if $F_{\rm obs}(N,z)$ is dominated by low--$N$ systems, and vice-versa.
Unfortunately, our evolution models $\rho_{\rm HI}(z)$ predict only the
first moment of $F_{\rm obs}(N,z)$, not the distribution itself.  However,
the DLA sample {\it does} tell us something about the full distribution; in
particular, the entire sample is well--fitted by a so--called gamma
distribution (Storrie-Lombardi, Irwin, \& McMahon 1996):
\begin{equation}
F_{\rm obs}(N,z) = F_\star
\left({N\over N_\star}\right)^{-\gamma}
\exp\left(-{N\over N_\star}\right),
\label{eq:gammadist}
\end{equation}
where $F_\star$ and $N_\star$ are functions of $z$ and $\gamma=$const.  If
we specialize to the case where $N_\star$ is a constant, then $F_\star$
contains all of the $z$--dependence, and is proportional to $dX/dz$ (i.e.,
all of the redshift dependence is due to the cosmological geometry).  In
this case, we may rewrite equation~(\ref{eq:fndn}) as
\begin{eqnarray}
n(z;\Omega_\Lambda,\eta,N_{\rm avg}) &\equiv&
\int_{N_{\rm min}}^\infty F_{\rm obs}(N,z)dN\nonumber\\
&=& \left({c\over{H_0m_{\rm H}}}\right)
\left(1\over{\N_{\rm avg}}\right)
\left[{{(1+z)^2g(z)}}\over{\sqrt{(1-\Omega_\Lambda)(1+z)^3 +
\Omega_\Lambda}}\right]{\ }\rho_{\rm s0}(1+\eta-f),\nonumber\\
\label{eq:fdn}
\end{eqnarray}
where
\begin{equation}
\N_{\rm avg} \equiv {{\int_{N_{\rm min}}^\infty F_{\rm obs}(N,z)NdN}
\over{\int_{N_{\rm min}}^\infty F_{\rm obs}(N,z)dN}} 
= N_\star \cdot {{\Gamma(2-\gamma,N_{\rm min}/N_\star)}\over
{\Gamma(1-\gamma,N_{\rm min}/N_\star)}}
\label{eq:navg}
\end{equation}
is the mean column density in the distribution, with
$\Gamma(a,x)\equiv\int_x^\infty t^{a-1}e^{-t}dt$ the incomplete gamma
function.  Note that equation~(\ref{eq:fdn}), with $\N_{\rm avg}$
independent of redshift, applies for {\it any} distribution $F_{\rm
obs}(N,z)$ which is separable into functions of $N$ and $z$.  The
assumption of separability is particularly appealing if only a small
fraction of the available HI is depleted from DLA galaxies during the
redshift range under consideration.  Suppressing the dependence on
$\Omega_\Lambda$, $\eta$, and $N_{\rm avg}$ for brevity, the quantity
$n(z)$ defined in equation~(\ref{eq:fdn}) denotes the redshift distribution
of absorbers; the number of systems observed with redshifts between $z$ and
$z+dz$ is $n(z)dz$.

For the moment, assume that we know the {\it shape} of $F_{\rm obs}(N,z)$
a priori, and hence that we know $\N_{\rm avg}$ from equation~(\ref{eq:navg}).
[We will address the question of how to incorporate properly our incomplete
knowledge of $F_{\rm obs}(N,z)$ from the data into our analysis in the next
section.]  Then we may construct a likelihood function from equation~(\ref{eq:fdn})
as follows.  Suppose we wish to compare the data to our prediction over
a range of redshifts ($z_{\rm min},z_{\rm max}$).  If we divide this range
into sufficiently small intervals $dz$, such that $n(z)dz\ll 1$, then there
will be at most one object in each such interval.  Then the probability of
finding no objects in a given interval at a redshift $z$ is given by the
Poisson distribution, $P(0)=\exp[-n(z)dz]$; similarly, the probability of
finding one object is $P(1)=n(z)dz\exp[-n(z)dz]$; and the probability of
finding more than one object is negligible,
$\sum_{i=2}^\infty P(i)\sim O\{[n(z)dz]^2\}\ll 1$.  Then the likelihood,
or conditional probability of obtaining a particular data set $D$ from a
distribution $n(z)$ predicted by~(\ref{eq:fdn}) is
\begin{equation}
P(D | \Omega_\Lambda, \eta, \N_{\rm avg}) =
\left\{\prod_{i=1}^{m_{\rm obs}} n(z_i)dz \exp[-n(z_i)dz]\right\}
\left\{\prod_{j=1}^{m_{\rm none}} \exp[-n(z_j)dz]\right\},
\label{eq:likea}
\end{equation}
where $m_{\rm obs}$ is the number of DLA systems observed, $z_i$ is the
redshift of the $i$--th system, $m_{\rm none}$ is the number of intervals
$dz$ in our range ($z_{\rm min},z_{\rm max}$) which have {\it no} DLAs
in them [note that $m_{\rm obs}+m_{\rm none} = (z_{\rm max}-z_{\rm min})/dz$],
$z_j$ is the redshift of the $j$th such interval, and we have suppressed the
dependence of $n(z)$ on the parameters $(\Omega_\Lambda,\eta,\N_{\rm avg})$
for brevity.  For $dz\ll 1$, equation~(\ref{eq:likea}) may be written as
\begin{equation}
P(D | \Omega_\Lambda, \eta, \N_{\rm avg}) =
\left[\prod_{i=1}^{m_{\rm obs}} n(z_i)dz\right]
\exp\left[-\int_{z_{\rm min}}^{z_{\rm max}} n(z)dz\right].
\label{eq:likeb}
\end{equation}
This probability clearly depends on the chosen size of the interval
$dz$, and shrinks to zero as $dz\rightarrow 0$.  This reflects the fact
that, as $dz$ shrinks, the number of possible distinct outcomes $D$ increases,
so the probability of obtaining any {\it particular} data set $D$ goes
to zero.  In the next section, we will use Bayesian inference to construct
confidence intervals from the likelihood function~(\ref{eq:likeb}).

\subsection{Computing Confidence Regions with Bayesian Inference}

Equation~(\ref{eq:likeb}) gives the probability of obtaining our DLA sample as a
random realization of the redshift distribution~(\ref{eq:fdn}), {\it assuming} the
truth of a particular cosmology $\Omega_\Lambda$ and evolution model $\eta$,
and assuming knowledge of the expected average column density $\N_{\rm avg}$
in the sample.  However, our evolution model for $\rho_{\rm HI}(z)$ does
not make any predictions about the value of $\N_{\rm avg}$; we must make use
of the DLA sample to extract information about the range of reasonable
values for $\N_{\rm avg}$.  In addition, the value of $\rho_{\rm s0}$ is
uncertain since it is related to the uncertain mass-to-light ratios of
present-day galaxies.  Both of these uncertainties in our knowledge may be
rigorously incorporated into our analysis if we use Bayesian inference.

Bayes' Theorem follows trivially from the axioms of probability and the
definition of conditional probability.  The theorem relates
$P(\Omega_\Lambda,\eta|D,\N_{\rm avg})$, the conditional probability
distribution for values of the model parameters given the observed data and
an assumed value of $\N_{\rm avg}$, to the likelihood 
$P(D|\Omega_\Lambda,\eta,\N_{\rm avg})$ [cf. eq.~(\ref{eq:likeb})].
The conditional probability of obtaining the observed data set as a
random realization of the model with particular parameter values is,
\begin{equation}
P(\Omega_\Lambda,\eta|D,\N_{\rm avg}) = A\cdot P(\Omega_\Lambda,\eta|\N_{\rm
avg}){\ }{\ } P(D|\Omega_\Lambda,\eta,\N_{\rm avg}).
\label{eq:bayesa}
\end{equation}
Here, $P(\Omega_\Lambda,\eta|\N_{\rm avg})$ is the {\it prior} probability
distribution for the two parameters, which, in the absence of any previous
data, we take to be uniform and independent of $\N_{\rm avg}$; and $A$ is a
normalization constant which ensures that $\int
P(\Omega_\Lambda,\eta|D,\N_{\rm avg})d\Omega_\Lambda d\eta=1$.  Note that
the value of $dz$ in equation~(\ref{eq:likeb}), which may be taken to be
arbitrarily small, is absorbed into $A$.  Unfortunately, our {\it a priori}
knowledge does not tell us the precise value of $\N_{\rm avg}$.  However,
we may include as an additional component of our model the assumption that
the data are well fitted by a gamma distribution~(\ref{eq:gammadist}).  By
fitting a functional form~(\ref{eq:gammadist}) to the data, we may obtain a
prior distribution $P(\N_{\rm avg})$ of reasonable values of $\N_{\rm
avg}$.  The procedure for obtaining $P(\N_{\rm avg})$ is identical to the
procedure we employ to get $P(\Omega_\Lambda,\eta | D,N_{\rm avg})$.  Note
that the gamma distribution is not the only plausible functional form to
fit to the data; however, it provides a better fit than a single power law
(Storrie-Lombardi, Irwin, \& McMahon 1996), since there is a significant
break in the observed slope of $F(N,z)$.  Any other reasonable
two-parameter distribution (e.g. a broken power law) will yield similar
results to the gamma distribution.

The sum rule of probability theory allows us to ``marginalize'' the parameter
$\N_{\rm avg}$ by integrating equation~(\ref{eq:bayesa}) over values of
$\N_{\rm avg}$, weighted by $P(\N_{\rm avg})$: 
\begin{equation}
P(\Omega_\Lambda,\eta|D) = A\cdot \int_0^\infty P(\N_{\rm avg}) {\
}P(D|\Omega_\Lambda,\eta,\N_{\rm avg}){\ }d\N_{\rm avg},
\label{eq:bayesb}
\end{equation}
where we have assumed that $P(\Omega_\Lambda,\eta|\N_{\rm avg})$ is
uniform, and have absorbed all constants into $A$ such that
$P(\Omega_\Lambda,\eta|D)$ is still normalized to unit area.
We may incorporate the uncertainty in $\rho_{\rm s0}$ in a
similar way.  As will be seen in \S 4.1, the uncertainty in $\rho_{\rm s0}$ is
approximately Gaussian.  Then, making the dependence on $\rho_{\rm s0}$
explicit, we finally have 
\begin{equation}
P(\Omega_\Lambda,\eta|D) = A\cdot \int_0^\infty d\N_{\rm
avg} {\ }\int_0^\infty d\rho_{\rm s0}{\ }
P(\N_{\rm avg}){\ }\exp\left[-{(\rho_{\rm s0}-\bar{\rho}_{\rm s0})^2\over{2\sigma^2}}
\right] P(D|\Omega_\Lambda,\eta,\N_{\rm avg},\rho_{\rm s0}),
\label{eq:bayesc}
\end{equation}
where once again the normalization has been absorbed into $A$.
We will use equation~(\ref{eq:bayesc}), substituting equations~(\ref{eq:fdn})
and~(\ref{eq:likeb})
for $P(D|\Omega_\Lambda,\eta,\N_{\rm avg},\rho_{\rm s0})$ and obtaining $P(\N_{\rm
avg})$ from a fit to the data, to compare the
data to the models and calculate confidence regions.

\section{Results: Application to the DLA Sample}

We include in our sample all DLA systems whose redshifts and HI column
densities (or, in a few cases, equivalent widths) have appeared in the
literature (Wolfe et al. 1986; Lanzetta 1991; Lanzetta 1995; Wolfe et al.
1995; Storrie-Lombardi, McMahon, \& Irwin 1996), for a total of 73 systems.
In the cases where no HI column density has been confirmed, we have
calculated it from the reported equivalent width using equation (3) in
Wolfe et al. (1986).  This is the same sample used by Storrie-Lombardi,
McMahon, \& Irwin 1996, with the addition of data from Wolfe et al. (1995).
In our analysis, we have concentrated on high redshifts ($z>2$), and hence
have used subsets of this sample.

\subsection{Calculating the Present-Day Stellar Density $\rho_{\rm s0}$}

Since our final results depend sensitively on the value and degree of
uncertainty in the local stellar density $\rho_{\rm s0}$, it is important
that we obtain the most accurate and precise possible estimate of this
quantity from the literature.  Previous studies (Wolfe et al. 1995;
Lanzetta et al. 1995; Storrie-Lombardi, McMahon, \& Irwin 1996) have
employed the value $\rho_{\rm s0}/\rho_{\rm c}=2.7\times 10^{-3\pm
0.18}h^{-1}$ (Gnedin \& Ostriker 1992), where $\rho_{\rm c}\equiv
3H_0^2/8\pi G$ is the present-day critical density, and the Hubble constant
is $H_0 = 100h$ km s$^{-1}$ Mpc$^{-1}$.  We make use of recent observations
to refine this estimate.

We compute $\rho_{\rm s0}$ by multiplying the local luminosity density of
galaxies by their mean stellar mass-to-light ratio.  Since the
mass-to-light ratio is in general correlated with the galaxy luminosity, we
have
\begin{equation}
\rho_{\rm s0} = \int_0^\infty \phi(L) \Upsilon(L) L dL,
\label{eq:rhos}
\end{equation}
where $\phi(L)$ is the galaxy luminosity function (LF) and $\Upsilon(L)$ is
the mass-to-light ratio in solar units.  We have used the luminosity
function for the NS112 sample of the Las Campanas redshift survey (Lin et
al. 1996), which represents the most precise determination of the local LF
to date.  We include all galaxy types, since recent identifications of DLA
galaxies (Le Brun et al. 1996) have indicated that DLA systems may be
associated with a wide range of morphological types.  Lin et al. (1996)
obtain a best-fit Schechter function with the parameters $M_\star =
-20.29\pm 0.02 + 5\log h$, $\alpha=0.70\pm 0.03$, and $\phi_\star =
(0.019\pm 0.001) h^3$ Mpc$^{-3}$, where the photometry was done in a band
very similar to the Cousins $R_c$ band.  Lin (1997, private communication)
points out that these Schechter parameters also provide a good fit for the
Gunn $r$-band luminosity function after a correction is made from isophotal
to total galaxy magnitudes.  Using a solar absolute magnitude of $r=4.83$
(Broeils, 1997, private communication), this yields an $r$-band luminosity
density of $j_r=(1.9\pm 0.1)\times 10^8 h{\ }L_\odot$ Mpc$^{-3}$.  For the
mass-to-light ratio $\Upsilon(L)$, we use the relation measured by Broeils
\& Courteau (1996) in the Gunn $r$-band for the disks of Sbc-type galaxies:
$\Upsilon(L)=[5.8(L/10^{10}L_\odot)^{0.24}\pm 1]h$, where the uncertainty
is the $1\sigma$ deviation and the distribution of observed values is
approximately Gaussian.  Thus, the uncertainty in $\rho_{\rm s0}$ is
dominated by the Gaussian uncertainty in $\Upsilon(L)$.  We include all
galaxy types in our calculation of $\rho_{\rm s0}$, with the assumptions that
spiral galaxies are well described by maximal--disk models, and that spiral
and elliptical galaxies have the same stellar mass-to-light ratio.
We take the mass-to-light ratio for Sbc galaxies as typical for all
galaxy types of the same luminosity.  Our approach overestimates the stellar
mass-to-light ratio somewhat if, as some studies suggest (Rix et al. 1997),
the dark matter accounts for a significant fraction of the total mass in the
inner regions of galaxies.  Using equation~(\ref{eq:rhos}), we obtain
\begin{equation}
{\rho_{\rm s0}\over{\rho_{\rm c}}} = (4.0 \pm 1.0) \times 10^{-3} h^{-0.48}.
\label{eq:rhosvalue}
\end{equation}
Based on the uncertainty in $\Upsilon(L)$, we assume that the values of
$\rho_{\rm s0}$ are normally distributed with a variance $\sigma=1.0\times
10^{-3} h^{-0.48}$.

\subsection{Effects of Dust}

The observed column density distribution of DLA systems, and hence the
inferred value of $\rho_{\rm HI}$, is affected by the presence of dust in
the DLA galaxies (Fall \& Pei 1993).  The dust obscures background QSOs,
and causes incompleteness in the DLA sample.
The presence of dust leads to an underestimate of $\rho_{\rm HI}$ from the
data; accounting for this fact will lead to better agreement between data
and predictions for high-$\Lambda$ cosmologies.  Fall \& Pei (1993) showed
that the true column density distribution is given by 
\begin{equation}
F_{\rm true}(N,z)
= F_{\rm obs}(N,z)\exp[\beta\tau(N,z)],
\label{eq:ftrue}
\end{equation}
where $\beta$ is the power-law slope at the bright end of the QSO luminosity function
[$\phi(L)\propto L^{-(\beta+1)}$], and $\tau$ is the extinction optical depth,
given by
\begin{equation}
\tau(N,z) = k(z)\left({N\over{10^{21}{\rm cm}^{-2}}}\right)
\xi\left({\lambda_{\rm e}\over{1+z}}\right).
\label{eq:taudef}
\end{equation}
Here, $k(z)=\rho_{\rm d}/\rho_{\rm g}$ is the dimensionless dust-to-gas
ratio, $\xi$ is the extinction curve, and $\lambda_{\rm e}$ is the
effective wavelength of the band of the QSO survey.  We have used
equations~(\ref{eq:ftrue}) and~(\ref{eq:taudef}), together with an
assumption that the observed distribution $F_{\rm obs}(N,z)$ may be fit by
a gamma distribution~(\ref{eq:gammadist}), and that $k(z)$ is proportional
to the metallicity, to estimate the correction to the expected value of
$\rho_{\rm HI}(z)$ due to dust obscuration.  Following Pei \& Fall (1995),
we assume that $k(0)=0.8$, $\beta=2$, and $\xi(\lambda)=\lambda_B/\lambda$
for QSO surveys in the $B$-band.  We assume that the metallicity at
redshifts $z=3\pm 1$ is approximately one-tenth of the solar value,
$Z(z=3\pm 1)=0.1Z_\odot$.  We find that the effect of dust is to reduce the
inferred value of $\rho_{\rm HI}$ by a factor of $\sim 1.5$.  Our results
in the next section will include this correction.

\subsection{Results of Bayesian Analysis}

We have applied equation~(\ref{eq:bayesc}), corrected for dust extinction,
to the DLA sample for two redshift
ranges, $2<z<4$ and $2.5<z<3.5$, allowing $0\le\Omega_\Lambda\le 1$, $-1\le
\eta\le 1.5$.  To compute the prior distribution $P(\N_{\rm avg})$, we first fit a
gamma-distribution~(\ref{eq:gammadist}) to the data, and use Bayesian
methods similar to those described in \S 3.2 to obtain a probability
distribution $P(\gamma,N_\star)$.  We fix $F_\star$ by requiring that the
integral of $F_{\rm obs}(N,z)$ over all column densities be equal to the
total number $m_{\rm obs}$ of DLAs observed in the narrow redshift interval
under consideration.  We then obtain a cumulative probability distribution
for $N_{\rm avg}$ by integrating numerically the probability
$P(\gamma,N_\star)$ inside contours of constant $\N_{\rm avg}$ in the
$(\gamma,N_\star)$ plane, given by equation~(\ref{eq:navg}).  We
differentiate this cumulative distribution numerically to obtain the
differential distribution $P(\N_{\rm avg})$.  Finally, we use
equations~(\ref{eq:fdn}) and~(\ref{eq:likeb}) to obtain the likelihood
function $P(D|\Omega_\Lambda,\eta,\N_{\rm avg},\rho_{\rm s0})$.
Substituting these results into equation~(\ref{eq:bayesc}) gives the
differential probability distribution $P(\Omega_\Lambda,\eta|D)$.
Confidence regions are obtained by integrating this distribution inside
contours of constant $P(\Omega_\Lambda,\eta|D)$.  Our results do not depend
strongly on the value of the Hubble constant; we assume $H_0=70$ km
s$^{-1}$ Mpc$^{-1}$.

In figure 1a, we show 68\%, 95\%, and 99\% confidence regions in the
$(\Omega_\Lambda,\eta)$ plane, for the redshift range $2<z<4$; figure 1b
shows the same for the redshift range $2.5<z<3.5$.  In both cases, we
assume that 25\% of the present-day mass density in stars had been
assembled by $z=2$ ($f=0.25$), consistent with the star formation rate of
Madau (1996).  The effect of the sample size is evident: the 99\%
confidence region is 20\% smaller for the larger sample (45 objects with
$2<z<4$) than for the smaller sample (17 objects with $2.5<z<3.5$).  Note
that, for the larger sample, the data are inconsistent with a net expulsion
of more than half of the gas from galaxies ($\eta>1$) at the 99\%
confidence level, and also rule out $\eta>0.5$ at 95\% confidence and
$\eta>0.1$ at 68\%, in all flat cosmologies.  One may compare this result
with the observational finding that the intracluster medium (ICM) in rich
clusters of galaxies contains an amount of iron which is approximately
equal to the amount found in the stellar component of galaxies in these
clusters (Renzini et al. 1993; Loewenstein \& Mushotzky 1996).  The natural
explanation is that the iron in the ICM was produced in supernova
explosions inside of galaxies, and was subsequently expelled into the ICM
by supernova--driven winds.  If galaxies were fully assembled by a redshift
$z\sim 4$, then they should have had twice as much baryonic material at
such redshifts than is observed locally.  Since our analysis excludes the
possibility that there was more than 1.5 (1.75) times the present--day
baryonic density at 95\% confidence for the larger (smaller) samples, it
suggests that (1) infall of material into existing galaxies or formation of
new galaxies has taken place since $z=2$; or (2) more than 25\% by mass of
the present-day stellar population formed by $z=2$.  These possibilities
will be discussed in more depth in \S 5.  Note also that for
high--$\Omega_\Lambda$ cosmologies, outflow models are strongly ruled out,
so it would be more difficult to reconcile the DLA data with the
intracluster iron observations in an $\Omega_\Lambda$-dominated universe.

If we wish to specialize to a particular cosmological model, we should
consider the one-dimensional probability distribution
$P(\eta|D,\Omega_\Lambda)$, since the sizes of the confidence intervals
decrease when the number of free parameters is reduced.  Figure 2a shows
the probability distribution for $\eta$, given $\Omega_\Lambda=0$ (the case
corresponding to the most conservative constraints on $\eta$).  We consider
the sample with $2<z<4$.  The solid curves correspond to different
assumptions about the amount of star formation that took place at $z>3\pm
1$: $f=0$, 0.25, and 0.5, from left to right.  Upper limits on $\eta$
derived from these curves are summarized in Table 1.  As $f$ increases, the
gas density observed in DLAs is supplemented by more and more stars, so the
data become more consistent with the closed--box model.  As can be seen
from Figure 2a, the data are consistent with modest to significant amounts
of infall or formation of new galaxies subsequent to $z\approx 3$.
Significant expulsion is less likely; for $f=25\%$ the data are
inconsistent with $\eta>0.42$ at 95\% confidence.  If one allows for half
of the present-day stellar population to form before $z=3$, then one
achieves marginal consistency with $\eta=0.92$ at 95\% confidence.
However, in this case the metallicity of DLA systems will significantly
exceed its observed value of $0.1 Z_\odot$ at $z\ga 2$ (Lu et al.  1996).
Figure 2b shows similar results for $\Omega_\Lambda=0.7$; here, no
significant amount of expulsion is viable.  Hence, it is unlikely that all
of the baryonic material seen today in galaxies and the ICM of rich
clusters was present in DLA galaxies at redshifts $z>2$.  The dotted curves
in Figure 2 are the results obtained by considering present-day galaxies +
ICM as a whole, with equal mass densities in stars and the ICM.  Hence, we
attempt to account for an amount ``$\rho_{\rm b,gal}(0)$''= $2\rho_{\rm
s0}$.  We assume that 25\% of the present-day mass in stars + enriched ICM
(that is, 0.5$\rho_{\rm s0}$) was present in the form of stars by $z=2$.
Clearly, only about half of this amount was present in DLA galaxies at
$z>2$ (see Fig. 2); the rest of it must have assembled into galaxies (and
some of it subsequently expelled) after $z=2$.  This possibility will be
discussed further in \S 5.

The constraints obtained with our method will improve as the catalog of DLA
systems grows.  The Sloan Digital Sky Survey, which is getting underway in
1997 (Gunn \& Knapp 1993; see also http://www.astro.princeton.edu/BBOOK),
will catalog $\sim 10^5$ quasars, at least an order of magnitude
more than the number discovered to date (Loveday 1996).
Spectroscopic follow-ups on this sample could increase the
DLA sample size by 1--2 orders of magnitude.  In Figure 3, we predict the
effect on our results of a more modest increase in the DLA sample size.
Figure 3a indicates how the probability distribution $P(\eta)$ changes when
the sample size $m_{\rm obs}$ with $2.5<z<3.5$ is increased by a factor of
2 or 5.  For illustrative purposes, we choose $\Omega_\Lambda=0$ and
$f=0.25$.  We take into account the observational uncertainty $\sigma$ in
the present--day stellar density $\rho_{\rm s0}$, and assume that the
additional DLAs have the same column density distribution as the existing
sample.  In this case it is a simple analytical matter to determine
$P(\eta)$ for a hypothetically larger data set.  The trend is obvious: as
more data are acquired, our measurement of $\eta$ becomes more precise, and
$P(\eta)$ becomes more sharply peaked.  Given our assumptions, the mean
value of $\eta$ will remain constant, but the confidence intervals will
shrink.  For the case shown in Figure 3a, the 95\% upper bound on $\eta$ is
0.32, 0.10, and -0.08, respectively, for $m_{\rm obs}=17$, 34, and 85.  In
reality, the mean value of $\eta$ will shift around as the column density
distribution becomes better known [i.e. $N_{\rm avg}$ in
equation~(\ref{eq:navg}) will not remain constant as observations improve],
but the width of $P(\eta)$ will still behave in the same way.  The
precision with which we can measure $\eta$ is limited by the 25\%
uncertainty in $\rho_{\rm s0}$.  Figure 3b demonstrates what would happen
if $\rho_{\rm s0}$ were known exactly.  In this case, the statistics are
purely Poisson and the width (and hence the height) of our normalized
distribution scales as $(m_{\rm obs}+1)^{1/2}$.

\section{Summary and Conclusions}

We have used the catalog of DLA systems to place constraints on the amount
of evolution in the baryonic content of galaxies and the value of the
cosmological constant.  We compared the gas density $\rho_{\rm g}$ at
redshift $z=3\pm 1$ to the present--day stellar mass density $\rho_{\rm
s0}$ in galaxies for a range of flat cosmologies with
$\Omega_\Lambda+\Omega_{\rm m}=1$.  Our underlying assumption is that cold
gas is required for star formation. We make use of the facts that DLA
systems dominate the HI content of the universe at redshifts $z>2$ and the
correction to their baryonic mass due to molecular gas is small,
$\la 20$\% (Ge \& Bechtold 1997).

We defined $\eta$ to be the net fraction of the baryonic content of local
galaxies which was expelled since $z\approx 3\pm 1$, and used Bayesian
inference to derive confidence regions in the ($\eta, \Omega_\Lambda$)
plane.  In all cosmologies we find that $\eta<0.4$ with at least 95\%
confidence, as long as only $<25\%$ of the current stellar population
formed before $z=3$. The most likely value of $\eta$ is negative (cf. Fig.
2), implying a net {\it increase} by several tens of percent in the
baryonic mass of galaxies since $z\approx 3$.  The inferred value of $\eta$
is more extreme for $\Omega_\Lambda$--dominated cosmologies.  On the other
hand, recent observations of high metal abundances in the intracluster
medium of rich clusters (Renzini et al. 1993; Loewenstein \& Mushotzky
1996) require that metal--rich gas be {\it expelled} from galaxies in an
amount approximately equal to the current mass in stars.  The possibility
that a dominant fraction of the present--day stellar population may have
already formed by $z=3$ and resulted early on in this expulsion, is ruled
out by the low metalicities ($0.01$--$0.1$ solar) observed in DLAs at
$z\ga3$ (Lu et al.  1996). Moreover, the intergalactic medium which later
accretes onto clusters of galaxies has a metallicity as low as $\sim
10^{-2}$ solar at $z\ga 2$ (Cowie~et~al.~1995; Tytler~et~al.~1995; Songaila
\& Cowie 1996; Cowie 1996), rather than the needed value of $\sim 0.3$
solar.  
Most of the required metal enrichment and star formation activity
must therefore have occurred at $z\la 2$.

The most likely explanation to the above discrepancy is that a significant
amount of gas had been assembled and partly expelled from galaxies after
$z=2$.  The increase in galactic mass could have been either in the form of
accretion onto existing galaxies or through the formation of new galaxies,
such as those responsible for the faint excess in deep galaxy counts (Lowenthal
et al. 1996, and references therein).  The likely value of $\eta$ of minus
several tenths (Fig. 2), implies that more than half the associated
baryonic mass was processed through galaxies after $z=2$. As an example,
let us assume that $\eta=-0.5$ at $z=3\pm 1$, and that 150\% of the current
baryonic mass of galaxies had assembled after $z=2$. This implies that the
total mass processed through galaxies is twice (150\%+50\%) their current
mass, as required by the observation that clusters contain twice the iron
locked up in stars (Renzini et al. 1993; Elbaz, Arnaud, \& Vangioni-Flam
1995; Lowenstein \& Mushotzky 1996).  Half of the processed mass was
converted into galactic stars and half expelled into the intergalactic
medium. If the expelled gas is $\sim 10\%$ of the intergalactic gas it
mixed with [assuming $\Omega_{\rm b}\sim 5\%$ and the value of $\rho_{\rm
s0}$ in Eq.~(\ref{eq:rhosvalue})], then it could have yielded the $\sim
0.3$ solar metallicity observed in clusters as long as its original
metallicity was a few times solar. 
%Such a metallicity is typical of giant ellipticals (O'Connell 1986).
The phase of massive metal enrichment must have occurred at $z\sim 0.5$--2
since the iron abundance in clusters shows little evolution at $z\la 0.5$
(Mushotzky \& Lowenstein 1997).  This inference could be tested
observationally through a dedicated search for enhanced star formation
activity and supernova rate at $z\sim 0.5$--2.

We have employed a Bayesian analysis which has the dual advantages of
taking the various observational uncertainties properly into account, and
making use of unbinned data.  The constraints obtained with our method will
improve as the size of the quasar sample increases (cf.  Fig. 3). In
particular, future spectroscopic observations of the $\sim 10^5$ quasars
cataloged by the Sloan Digital Sky Survey
(http://www.astro.princeton.edu/BBOOK), could increase the current sample
size of DLAs by 1--2 orders of magnitude, and improve our limits on the
amount of galactic evolution considerably.

\acknowledgements

We thank Tom Loredo for useful comments.
AL was supported in part by the NASA ATP
grant NAG5-3085 and the Harvard Milton fund. 

\vfil
\eject

\vfil 
\eject
\bigskip

\halign{
\tabskip=3em
#\hfil & \hfil#\hfil & \hfil#\hfil & \hfil#\hfil & \hfil#\hfil & \hfil#\hfil\cr
\multispan{5} \hfil \hfil \cr\cr
\noalign{ \hrule height .08em \vskip 2pt \hrule height .08em \vskip 5pt}
& Confidence & & Upper Limits on $\eta$ & & (Stars + ICM) \cr
$\Omega_\Lambda$ & Level & $f=0.00$ & $f=0.25$ & $f=0.50$ & $f=0.25$ \cr
\noalign{\vskip 5pt \hrule height .08em \vskip 5pt}
0.0  & 68\% & -0.30 & +0.18 & +0.72 & -0.40\cr
     & 95\% & -0.06 & +0.42 & +0.92 & -0.14\cr
     & 99\% & +0.20 & +0.66 & +1.10 & +0.20\cr
0.7  & 68\% & -0.60 & -0.32 & +0.06 & -0.54\cr
     & 95\% & -0.36 & -0.06 & +0.30 & -0.40\cr
     & 99\% & -0.10 & +0.18 & +0.54 & -0.20\cr
\noalign{\vskip 5pt \hrule height .08em \vskip 5pt}
}
\noindent {\bf Table 1:} Upper limits on $\eta$ for two cosmologies 
and different fractions $f\equiv\rho_{\rm s}(3)/\rho_{\rm s}(0)$ of
stars formed at redshifts $z=3\pm 1$.  The last column includes the
mass density in the intracluster medium (ICM), assuming 25\% of the
total (stars+ICM) was in the form of stars at $z=3\pm 1$.
See Figure 2 for the differential probability distributions $P(\eta)$
corresponding to columns 3--6.

\begin{figure}
\plotone{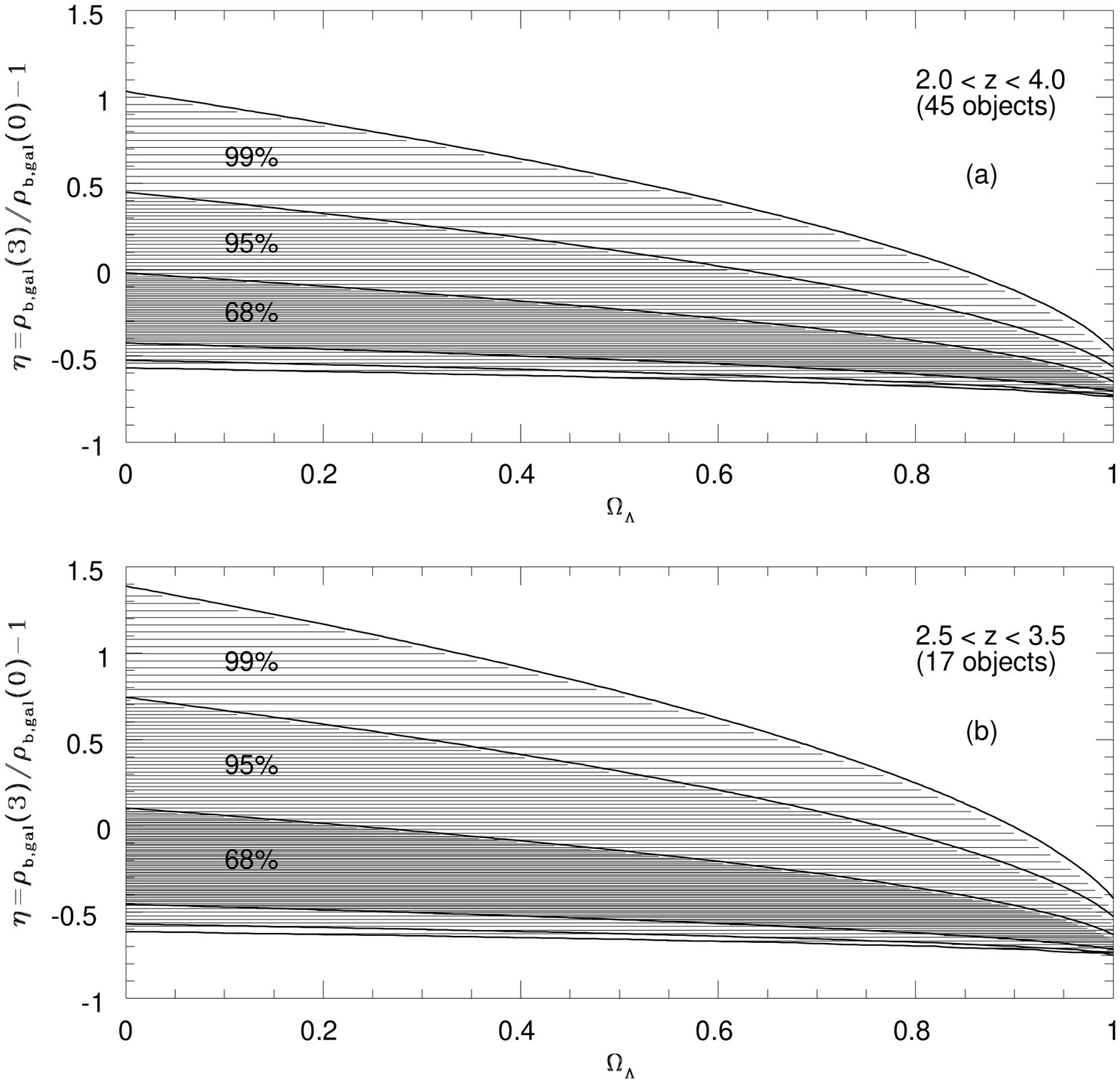}
\caption{Confidence regions in the $(\Omega_\Lambda,\eta)$ plane, computed
with dust obscuration.  Results are shown for two redshift intervals
over which the HI density is calculated:
(a) $2<z<4$; (b) $2.5<z<3.5$.}
\end{figure}

\begin{figure}
\plotone{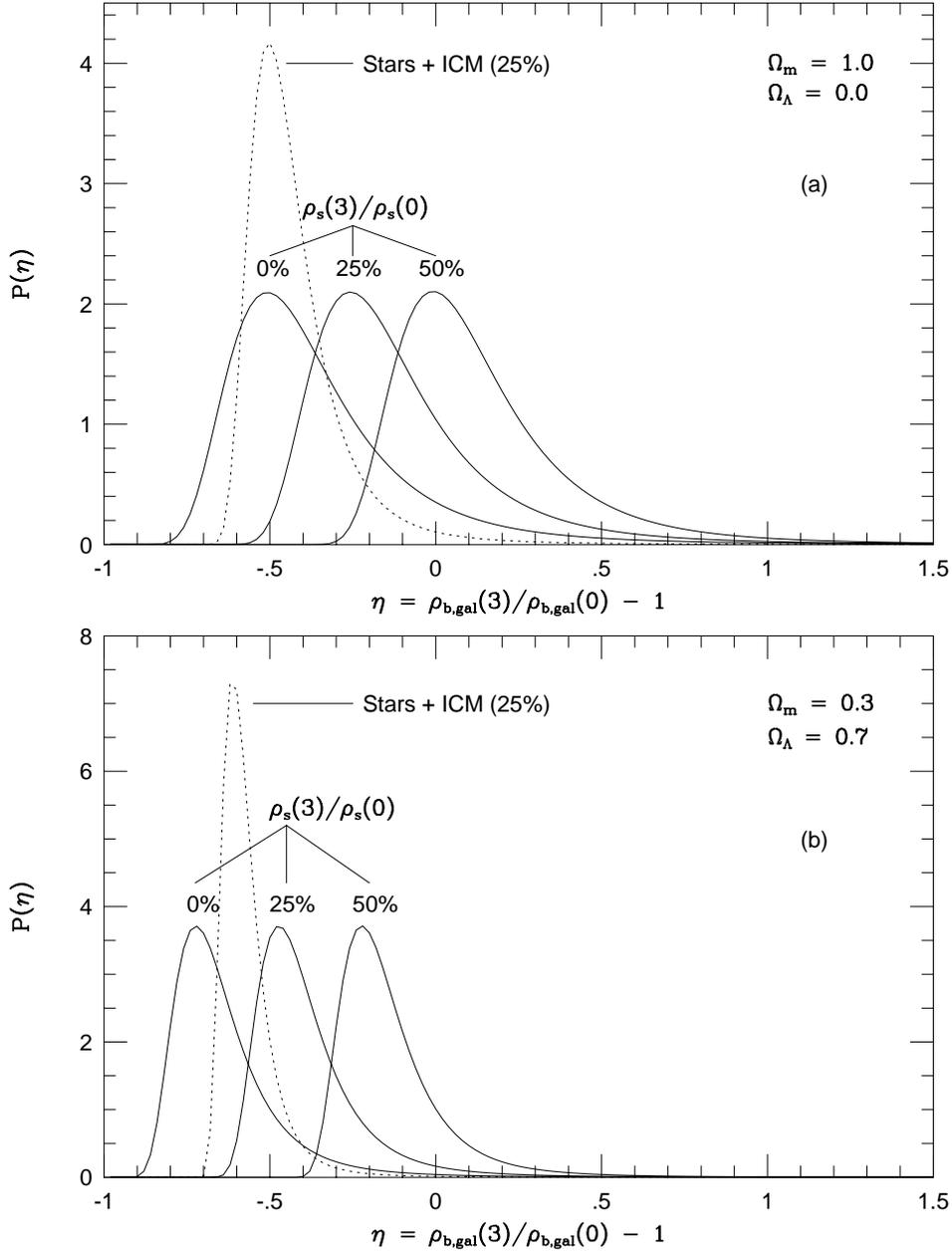}
\caption{Probability distributions $P(\eta)$, for (a) $\Omega_\Lambda$=0;
(b) $\Omega_\Lambda=0.7$.  We use the $2<z<4$ subsample. {\it Solid
curves}: results for three different amounts of star formation before
$z=3\pm 1$, namely 0\%, 25\%, and 50\% of the present--day stellar density.
{\it Dotted curves}: including the mass density in the intracluster medium
(ICM), assuming 25\% of the total (stars+ICM) was in the form of stars at
$z=3\pm 1$.}
\end{figure}

\begin{figure}
\plotone{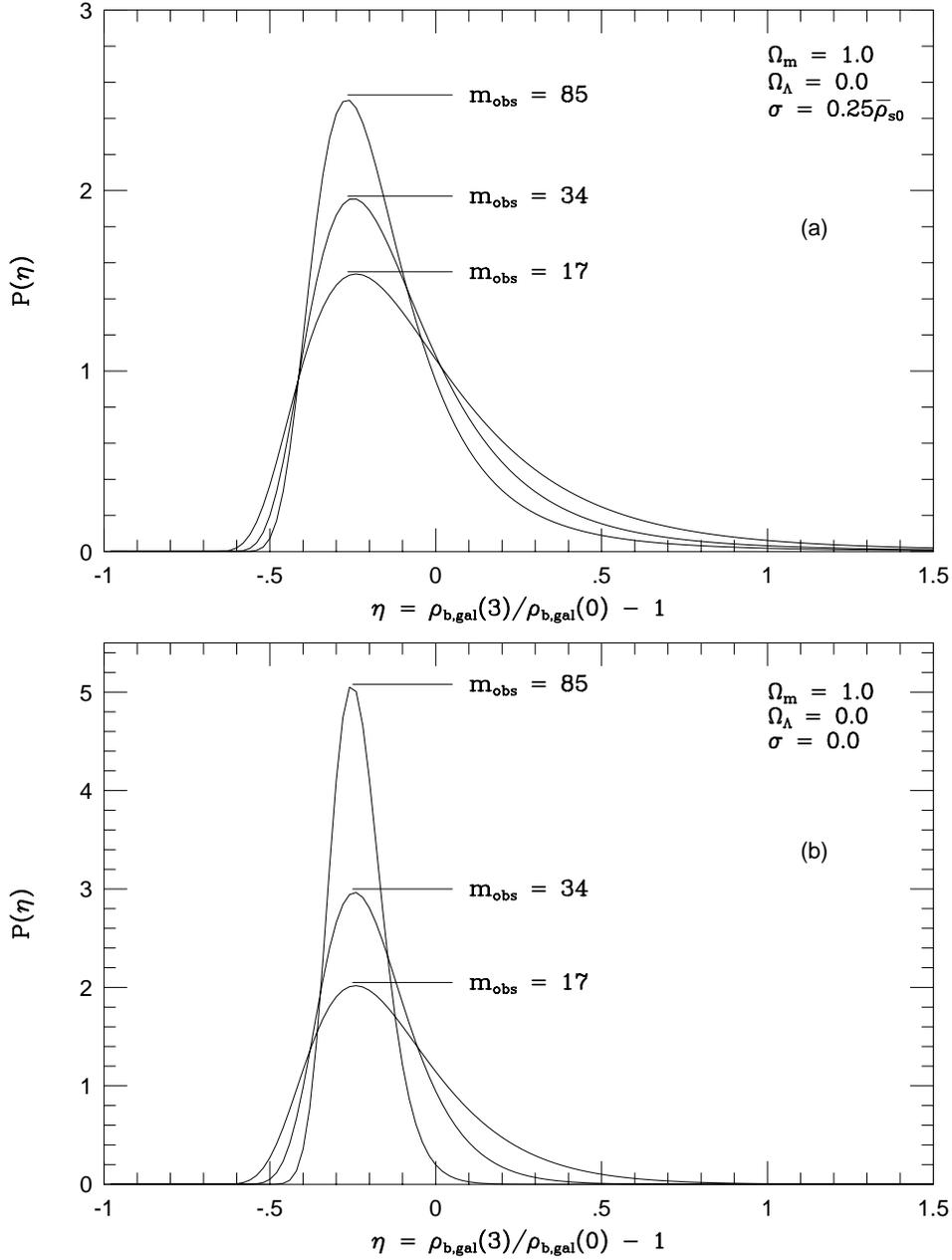}
\caption{Effect of increasing the DLA sample size $m_{\rm obs}$,
(a) including the 25\% observational uncertainty in the present--day
stellar density $\rho_{\rm s0}$; (b) assuming we know $\rho_{\rm s0}$ exactly.
We use the $2.5<z<3.5$ subsample.}
\end{figure}

\end{document}